\documentclass[a4paper,12pt]{article}

\newcommand{\myTitle}{Understanding news story chains using information retrieval and network clustering techniques\thanks{Thanks to colleagues from the Political Methodology Specialist Group of the Political Studies Association who commented on an earlier presentation of this method at their 2017 meeting.}}
\usepackage[a4paper,left=3.25cm, right=3.25cm, bottom=3cm, top=3cm]{geometry}
\usepackage{graphicx} 
\usepackage[british]{babel}					
\usepackage{csquotes}
\usepackage{multirow} 
\usepackage{booktabs} 
\usepackage{dcolumn}
\usepackage[titles]{tocloft}  
\usepackage[printonlyused]{acronym} 
\usepackage[dvipsnames]{xcolor}

\usepackage[natbib=true,style=authoryear-comp,bibencoding=utf8,date=year,maxbibnames
=20,urldate=short,backend=biber]{biblatex}

\DeclareSourcemap{
     \maps[datatype=bibtex]{
         \map{
             \step[fieldsource=url,final]
             \step[fieldset=doi,null]
             \step[fieldset=issn,null]
         }
        \map[overwrite]{
            \step[fieldsource=url,
                  match=\regexp{^https?://(dx\.)?doi\.org/.*},
                  fieldset=urldate,
                  null]
            \step[fieldsource=url,
                  match=\regexp{^https?://doi\.acm\.org/.*},
                  fieldset=urldate,
                  null]
        }
        \map{
            \step[fieldsource=note,
                  match=\regexp{.*DOI.*},
                  final]
            \step[fieldsource=note,
                  fieldtarget=doi]
            \step[fieldsource=doi,
                  match=\regexp{.*\{?DOI\}?: (.*)},
                  replace=\regexp{$1}]
        }
     }
}

\DeclareTextFontCommand{\textcsc}{}

\DeclareTextFontCommand{\textlf}{\addfontfeatures{Numbers=Lining}}

\usepackage{tikz}
\usetikzlibrary{shapes.geometric, arrows, positioning}
\tikzstyle{startstop} = [rectangle, rounded corners, minimum width=3cm, text width=3cm, minimum height=1cm,text centered, draw=black, fill=Maroon!30]
\tikzstyle{io} = [trapezium, trapezium left angle=70, trapezium right angle=110, minimum width=3.5cm, text width=2cm, minimum height=1cm, text centered, draw=black, fill=blue!20] 
\tikzstyle{process} = [rectangle, minimum width=3cm, minimum height=1cm, text centered, text width=3cm, draw=black, fill=orange!30]
\tikzstyle{decision} = [diamond, aspect=2, minimum width=4cm, minimum height=1cm, text centered, draw=black, fill=green!30]
\tikzstyle{arrow} = [thick,->,>=stealth]

\usepackage{tabularx} 

\newcommand{\listModelname}{List of regression models}
\newlistof{Model}{mdl}{\listModelname}

\begin{document}


\title{\myTitle}

\author{Tom Nicholls$^1$ \and Jonathan Bright$^2$}
\date{%
    $^1$Reuters Institute for the Study of Journalism, University of Oxford\\%
    $^2$Oxford Internet Institute, University of Oxford\\[2ex]%
    \today
}

\maketitle

\begin{abstract}
Content analysis of news stories (whether manual or automatic) is a cornerstone of the communication studies field. However, much research is conducted at the level of individual news articles, despite the fact that news events (especially significant ones) are frequently presented as ``stories'' by news outlets: chains of connected articles covering the same event from different angles. These stories are theoretically highly important in terms of increasing public recall of news items and enhancing the agenda-setting power of the press. Yet thus far, the field has lacked an efficient method for detecting groups of articles which form stories in a way that enables their analysis. 

In this work, we present a novel, automated method for identifying linked news stories from within a corpus of articles. This method makes use of techniques drawn from the field of information retrieval to identify textual closeness of pairs of articles, and then clustering techniques taken from the field of network analysis to group these articles into stories. We demonstrate the application of the method to a corpus of 61,864 articles, and show how it can efficiently identify valid story clusters within the corpus. We use the results to make observations about the prevalence and dynamics of stories within the UK news media, showing that more than 50\% of news production takes place within stories. 

\end{abstract}

\section{Introduction}

Content analysis of published news is one of the most common techniques in studies of mass communication and journalism. This analysis, which is frequently supported by large scale manual coding efforts (and more recently by automated techniques), has underpinned investigations into many of the core theories in the field such as news values \citep{Harcup2001}, news agendas and agenda setting \citep{Iyengar1993}, news diffusion and sharing \citep{Bright2016}, gatekeeping and editorial decision making processes \citep{bright_life_2014}, and news readership dynamics \citep{Graber1988}, to give but a few examples.

A common simplification that is, to our knowledge, found in the vast majority of these studies is the coding of content at the level of the individual news article (often considering articles which were published on the ``front page'' of the newspaper or online news portal in question). However, this marginalises a second potential level of observation, which we call here news ``stories'': collections of news articles which may approach the same subject from different angles, or consist of an initial piece and then follow-up reporting on a news event of continuing interest. These story groupings are theoretically significant because they have more potential impact than individual news articles, and indeed \textit{prima facie} evidence suggests that the news media itself reserves them for what are perceived to be the most important or significant issues of the day. However, research on news stories has been limited thus far largely because of the technical difficulties of observing and measuring these stories. Many content analysis efforts focus on a handful of ``constructed weeks'' of data, within which few stories may actually be found. If analysis is expanded to cover a greater range of observations, then actually identifying stories starts to become highly time intensive, as their accurate identification involves comparing all possible pairs of articles within a dataset.  

In this paper, we present a novel method for the automatic detection of news stories within news article data which resolves these problems. The method draws from two distinct fields: information retrieval for measuring textual similarity between articles, and network analysis for clustering articles into story groups. We also employ a moving window to reduce the computational complexity of the operation so that the method itself can be applied to a corpus of articles of any size. 

The rest of this paper is structured in the following way. In part 1, we discuss the concept of news stories in more detail, highlighting the limited amount of existing literature on the subject, and showcasing why stories are theoretically highly important even if in practice they remain understudied. We also explore in more detail the practical difficulties involved in identifying story clusters with manual content analysis techniques. 

In part 2, we introduce our method, positioning it as part of the wider field of automatic content analysis within the social sciences. We show how information retrieval and network analysis approaches offer a useful complement to existing techniques based on dictionaries, supervised learning and unsupervised topic detection. We also outline in detail the steps of the method. 

In part 3, we apply the method to a news corpus containing 61,864 articles, and perform a validation on the basis of a smaller dataset of hand-coded story data. The results show that the method performs well, and also allow us to draw out some first order descriptive insights about the prevalence of stories within the UK news media, and the dynamics by which they emerge and dissolve. 

\section{Stories in the News Media: Definition, Theory and Potential Impact}

In this paper, we define news ``stories'' as events which receive repeated coverage in the news media through a linked chain of news articles. These news stories are conceptually distinct from news ``topics'' or ``issues'', which of course do naturally receive repeated coverage, though each story will be associated with a topic. For example, ``health'' would be an example of a general news topic which is typically present in most news code books for content analysis\footnote{See e.g. the Policy Agendas news codebook, available from: \url{http://www.policyagendas.org.uk/}}, whereas the Affordable Care Act would be an example of a particular news story within the health topic area. 

While not always explicit, the importance of the distinction between an ongoing news story and a one-off news article is a feature of a wide variety of strands of research in journalism and communication studies. These works have highlighted two main reasons why they are theoretically important. Firstly, in research on news values and journalism, the potential for follow-up reporting is sometimes listed as a motivating factor for publishing an initial article. For example, spectacular crimes have been known to lend themselves to repeated coverage \citep{Peelo2006}, something that could be a motivating factor in their initial publication. Which events are suitable for follow-up is something that can evolve over time: research has highlighted, for example, that particular types of crime coverage have served as ``prototypes'' that are then repeated in later stories \citep{Brosius1995}. 

Once an initial article has been published, follow-up pieces also seem to become more likely, even if they could not be foreseen at the time of initial publication. Indeed, Harcup and O'Neill define ``follow-up'' as a news value in and of itself \citep{Harcup2001}, while Vasterman has claimed that the news ``threshold'' for follow-up articles may be lower than that for initial articles \citep[p.514]{vasterman_media-hype:_2005}. A variety of authors have highlighted of course that those working in the news media may actively ``manufacture'' fresh angles for follow up stories on events they consider particularly worthy of coverage \citep[p.7]{Chadwick2011}. All of the above points to the first major reason why news stories are important: because they form an important part of how journalists and editors think about the news, and because they may drive publishing decisions. 

Secondly, a small body of empirical work in communication studies has focussed on the idea that news production appears to have, potentially, two different ``modes'' \citep{boydstun_two_2014}: a mode of normal production, and a mode characterised by intense focus on a single issue, where large amounts of coverage are dedicated to a single story. These moments of focus have been called, variously, ``media hypes'', ``news waves'' and ``media storms'' \citep{vasterman_media-hype:_2005,wien_anatomy_2009,boydstun_two_2014,waldherr_emergence_2014}, but all of these terms capture the same basic premise: the news media as a whole dedicate themselves to an in-depth examination of a particular current event, with multiple follow up pieces and different angles explored. Major terrorist attacks \citep{Entman2003}, catastrophes such as the Challenger shuttle explosion \citep{Riffe1989}, or political scandals \citep{Chadwick2011} present examples of such media storms. 

These storms have significant potential consequences. Most obviously, pieces of news which become media storms are more likely to reach widespread public attention: repeated publishing increases the likelihood that people are exposed to news, whilst those people who read multiple articles on the same topic are likely to receive a signal about its importance. Furthermore, even in the age of online journalism, news production is still largely a ``zero sum'' game \citep{zhu_issue_1992}, whereby increase in attention to one topic or event must imply decrease in attention to another. Boydstun et al. find that just over 11\% of news coverage is, on average, attributable to very large ``mega stories'', that last for around 15 days on average \citep[p.520]{boydstun_two_2014}. Most seriously, perhaps, Vasterman has argued forcefully that media storms can often inflate a given news event beyond any objective measure of its actual importance and significance, as the news media start to involve themselves in a self-referential cycle which is detached from other ongoing events, such that ``even the most trivial details [about the event] can become the most important news fact of that day'' \citep{vasterman_media-hype:_2005}[509]. These moments are often when the news media is perceived as having the most influence on things like the political agenda \citep{VanAelst2011}[303]. All of the above points to a second major reason why news stories are important: because they may be occur when the news media enter this second ``mode'' of news production, where it has particular influence over political agendas.    

\section{Detecting news stories in media content}

Despite the theoretical importance of news story chains, in practice they have attracted relatively little empirical research. Indeed, the vast majority of work on news content production and news events takes place at the level of the individual article. The main reason for this, we believe, is that the identification of groups of linked news stories is prohibitively expensive in terms of researcher time. Determining whether two articles address the same topic requires a researcher to perform a pairwise comparison between the two articles. The number of pairwise comparisons required to exhaustively evaluate a given dataset of $n$ articles is $\frac{n(n-1)}{2}$. Even for a small dataset of 100 news articles (less than the amount produced on a typical news website in a typical day), fully 4,950 separate comparisons would have to be performed to detect all possible stories. Of course, with such a small baseline set, a researcher might simply be able to scan all stories at once and quickly pick out article groups, without going through each individual pair. But as the number of articles grows this strategy would be increasingly likely to generate errors. More formally, this type of pairwise comparison can be said to take ``quadratic time'': i.e. the length of time required grows with the square of the number of input articles (often represented as $O(n^2)$ in computer science). 

This difficulty of producing wide ranging and systematic story datasets is reflected by the methodological sacrifices made by research work up until now. Of the work on news storms referenced above, Waldherr et al. use simulated data, whilst both Vasterman and Wien \& Elmelund-Præstekær cherry pick news stories which are previously known to be important, which is a reasonable strategy for making initial observations but undermines potential generalisability. Boydstun et al. are the closest to being able to execute a fully quantitative approach, however their methodology relies both on a hand coded dataset of over 50,000 articles separated into more than 230 categories (a coding effort that would be extremely time consuming to reproduce) and also a heuristic method for detecting stories based on a sudden increase in the coverage of particular topics (a method that appears well suited to capturing some stories but that is very unlikely to capture all of them). 

Recent advances in automatic content analysis present the opportunity to resolve some of the above problems. Automatic content analysis is a growth area in communications research, inspired by advances in the fields of both corpus linguistics and machine learning. In their overview article on the subject \citet{grimmer_text_2013} identify three main types of automatic content analysis technique which have so far been applied: dictionary methods, supervised methods, and automated (unsupervised) clustering. In this section we briefly review each of these techniques, explaining why these standard approaches are inappropriate for our specific problem of extracting stories, even if they can perform well in labelling broader topics.

The dictionary approach is probably the most widely used current approach to automatic content analysis, and also in some senses the simplest. Dictionary approaches involve developing lists of key words which relate to particular topics of interest: for example, the word ``doctor'' might be associated with a health topic, whilst the word ``budget'' might be associated with economic topics. Classification decisions are based on the appearance of keywords in a given document, typically weighted by the frequency of appearance. Dictionary approaches are frequently applied by researchers interested in one specific issue or topic. For example, keyword searches could be employed to find news articles, political texts or other documents pertaining to one specific event of interest. In these instances, assuming the keyword list is well developed, the approach can be useful, especially if results are then checked by hand (though of course establishing how many documents were not found by such a technique is difficult). The major challenge in dictionary approaches is the development of the keyword lists, which need to be created by hand. Beyond the raw intuition of the researcher, there is little way of systematically developing these lists (which is one reason why these methods are increasingly criticised).

This difficulty has stimulated the adoption of supervised machine learning methods. These methods take a step on from dictionary approaches by systematizing the development of selection criteria. Rather than developing the lists themselves, human coders separate a training set of documents into pre-existing categories of interest. A computer program is then used to examine the documents in each category, extracting features (which are typically words but could also be phrases, punctuation, document length, or anything else for that matter) which are the most ``informative'' for the purposes of classification, which means that they appear frequently in the document of interest and infrequently everywhere else. A variety of techniques exist for both determining which features are most informative and then making classifications on the basis of these features. Supervised machine learning is increasingly popular because it is systematic, easy to validate, and a number of effective off-the-shelf methods are available.

However, while both techniques are useful if applied in the right circumstances, neither of them are suitable to the task of extracting large numbers of \textit{stories} from a corpus of news articles. This is because both dictionary approaches and supervised machine learning techniques require definition of categories of interest before classification can be performed: in the case of dictionary approaches these categories are required to develop keyword lists; in the case of supervised learning they are required to allow a training set of documents to be coded. These steps are themselves only feasible when the number of categories is reasonably limited: otherwise the human labour involved becomes prohibitive.     

Unsupervised clustering methods are an alternative approach to classification which has been applied less in the social sciences, though there is increasing interest in the use of topic modelling using unsupervised approaches \citep{roberts_structural_2014,guo_big_2016}. Rather than assuming categories before the classification commences, these approaches aim to extract structure purely from the observation of the data (using a logic similar to Principal Components Analysis). Based on a given set of textual features, the aim is to separate documents into clusters which are as homogeneous as possible, whilst maximising differences between clusters. A key feature in unsupervised clustering is the use of a distance measure \citep[p.321]{manning_introduction_2008}: after coding documents as sets of features, the distance between them is established, and this allows a determination to be made about whether they form part of the same cluster. 

However, whilst unsupervised methods allow us to work with a potentially greater number of topics, the standard approaches also suffer from limitations for the particular use case we have imagined here. First, they typically require the researcher to specify the number of categories ($k$) to be used in advance. It is difficult to do this in a systematic and principled way; researchers typically fit a large number of models across a wide range of values for $k$ and select between them post-hoc, decreasing researcher degrees of freedom. Second, the number of categories is usually reasonably limited. This is because after clustering takes place, researchers typically need to interpret the substantive significance of the categories themselves. Hence such methods are difficult to apply directly to the task of extracting an unknown (but potentially very large) quantity of stories from a corpus of news articles. 

As all the automatic analysis methods we have identified above are largely unsuited to our task, in this paper we offer a new approach to the detection of story clusters in news corpora, based on techniques drawn from the field of information retrieval and network analysis. The method is designed to be applied to a corpus of news articles of any size. It involves two steps. First, we use information retrieval methods to measure the pairwise similarity between different news articles in the corpus. Second, we conceptualise these similarity measures as connections in a network of articles, and then use clustering techniques from network analysis to detect related stories. We will describe each of these steps in turn here. 

\subsection{Calculating pairwise similarity with information retrieval approaches}

Information Retrieval [IR] is a set of approaches, drawing on the tools of computer science, information science, and corpus linguistics, for accurately locating small amounts of relevant information in large data sources \citep{manning_introduction_2008}. Its most prominent modern application is in search engines, which are the archetypal IR system: from a given query, they need to identify and retrieve the most appropriate documents from a vast range of possibilities. Some of the tools used in IR overlap with the more general approaches already used by social scientists using large-scale text analysis. However, IR contrasts with more general supervised classification approaches in that the relevant unit of interest is often a very small proportion of the whole, rather than a small number of larger groups. IR techniques are typical where the key problem is to select a few relevant documents for a given query, rather than to partition all the available documents into a known number of groups. These techniques at heart look at the content and structure of underlying documents, indexing the information contained therein to allow answers to be given to arbitrary queries. IR approaches are useful in our context because they provide a number of ways of thinking about the extent to which two documents are ``similar''. This similarity is at the heart of what it means to be a chain of articles all related to the same overall news event. 

We employ two IR techniques to develop our pairwise similarity measures. First, we identify and score the most distinctive words in each article compared to the corpus as a whole by relative frequency, allowing documents to be labelled with key distinctive terms. Each term $t$ is scored based on its frequency $f$ in the document $d$ and in the corpus of all text:

\begin{equation}
\label{eq:kw}
\displaystyle kwscore(t) = \frac{f_{t,d}}{f_{t}}
\end{equation}

In our implementation, the words in each article are scored using Equation~\ref{eq:kw} and ranked, with the top 100 most informative terms for each article recorded\footnote{The list is limited to terms with a $kwscore$ of greater than 100, to avoid short articles generating spurious non-keywords.}. This allows us to calculate a keyword similarity score for each pair of articles, which is simply the proportion of keywords which are common to both articles' top 100 list.

This scoring method selects strongly for the most unusual words in a given document. Although this is fairly naïve in IR terms, it is theoretically very suitable for news clustering. The intuition here is that news stories are about something concrete: a place, a person, or an event. By finding the most unusual terms in each article compared to the full output of the parent news source, it is possible to extract with some specificity the most distinctive words in each article. If articles share keywords, they are presumptively about the same subject. For this reason, we do not discard rare stopwords, as is conventional -- here they are central to the theoretical justification for the method \citep[see, more generally,][p.273, for the importance of the research question in choice of approach]{grimmer_text_2013}.

For the second part of the pairwise similarity measurement, we use the BM25F scoring algorithm to select related articles. BM25F is an example of the class of \textit{scoring rules}, which are used in information retrieval applications to identify which documents amongst many candidates are most relevant.

BM25F is a standard general purpose scoring algorithm. It is a development of Okapi BM25, which handles (as in our case) documents with multiple separate fields (body and title). Unlike the keyword approach identified above, which simply selects the most important terms in an article, BM25F scores documents in relation to a query. In a search engine context, this would be the text entered by the user; when attempting to find similar documents, it uses the content of the document being matched against as the query key\footnote{This is a simplification and elides the intermediate step of query expansion, but this can be automatically handled; for this work we used Bo1, one of the standard Bose-Einstein query expansion models, to create the final BM25F query from the text of the article being matched.}

The following BM25F equations are drawn from \citet{perez-iglesias_integrating_2009}:

\begin{equation}
\label{eq:bm25fscore}
 \displaystyle \textit{BM25Fscore(q,d)} = \sum_{\textit{t in q}} \log{\left(\frac{N-df(t)+0.5}{df(t)+0.5}\right)} \cdot \frac{w(t,d)}{k_1 + w(t,d)}
\end{equation}

Where $q$ is a given query and $d$ a document, $N$ is the number of documents in the collection and $df$ is the number of documents in which the term $t$ appears. 

The accumulated weight of a term over all fields $w(t,d)$ is calculated as follows:

\begin{equation}
\label{eq:bm25fwt}
\displaystyle \textit{w(t,d)} = \sum_{\textit{c in d}} \frac{occurs_{t,c}^d \cdot boost_c}{((1-b_c)+b_c \cdot \frac{l_c}{avl_c})}
\end{equation}
Where $l_c$ and $avl_{c}$ is the average length for the field $c$, 
and $boost_c$ is the boost factor applied to field $c$\footnote{We use $boost_{title}=2$ and $boost_{body}=1$ to give a modest increase in the importance of words in the title.}. $k_1$ and $b$ are free parameters (with $b$ free to vary between fields), which can be empirically selected to best improve the results of the subsequent objective function, or left at reasonable default values (such as $k_1 \approx 1.2, b \approx 1$). 

Our pairwise similarity measure is the mean of these these two measures: the proportion of keywords in common, and the BM25F score between the two articles. As we have remarked above, pairwise similarity calculations take polynomial time to produce (often denoted as $O(n^2)$). Although the automatic nature of the calculations means that time taken is less important than it would be for manual operations, it nevertheless can be a significant impediment to research work if run-time starts to be calculated in days or weeks\footnote{The fairly simple Python implementation of BM25F scoring used by the authors, for example, requires a few seconds of CPU time per comparison, which would quickly become infeasible on large datasets.}. A simple way of reducing this complexity is to only conduct pairwise calculations within a moving window. News stories are, in some sense, intrinsically time-bound entities, with articles being released in close proximity. Hence by restricting the time window within which comparisons are made, we can restrict the overall run time of the method. 

The disadvantage of this approach is that it will, obviously, not identify stories whose publication arc takes place outside the time window in question. For example, a crime might lead to a prosecution weeks after its original development, and then a court case which takes place months or even years after that. Alternative techniques are therefore available for general-purpose complexity reduction in IR which require less subject knowledge but a stronger view of the initial query. One is use of the Boolean IR model, which simply uses unscored Boolean query matching/not matching (via computationally cheap lookup techniques such as hash tables of features) to select candidate texts to analyse with more sophisticated and expensive IR scoring methods \citep[see][Ch.1]{manning_introduction_2008}.   

\subsection{Construction and partitioning of a similarity network}

Having constructed some kind of metric-based way of scoring documents in response to a query, it is necessary to identify which documents are considered `matching'. The two main approaches taken in IR are rank ordering and using a score cutoff. The first is familiar from search engine use: those pages the system has identified as `most relevant' are shown first and further pages of responses can be fetched until the user is satisfied with one of the pages retrieved or abandons the search. The second is more common in partitioning problems: having identified that certain documents are partially similar (or, using more probabilistic methods, have a given probability of being in the same group) then those under a certain threshold can be discarded, and those above presented as part of a set to the user.

The threshold approach works well in a pairwise context (are these two items part of the same group or not?) but less well where there are multiple groups into which documents can be placed, and it is undesirable to allow cutoff scoring to potentially place documents into either zero or two plus groups each.

We hence use an alternative approach, of converting a matrix of pairwise similarity metrics into a similarity network, then applying network partitioning tools to supply the boundaries. Networks inherently arise in the context of object-by-object comparison, as these similarity judgments have the natural interpretation of relating how far apart two documents are in some sense. As these comparisons are made for increasing numbers of document pairs, the table of scores becomes equivalent to an unrolled network, which is then tractable by standard methods for analysing and partitioning networks.

This problem, of taking a network-based representation of data, and simplifying and grouping the network nodes into groups is called community detection. There are a large number of community detection algorithms available, based on the optimisation of the modularity property of the graph and otherwise \citep[see][for a discussion]{lancichinetti_community_2009}. We make use in particular of the Infomap method \citep{rosvall_map_2009}. This method aims to detect clusters in the network by modelling a random walk on the network, and by optimizing a quality function based on compression of the information contained within the network by minimising the description length of the random walk \citep[p.4]{lancichinetti_community_2009}. It has the practical benefit for this application that it appropriately handles network links which are both weighted and directed, and that the partitions completed are fully hierarchical. Unlike many unsupervised approaches, no prior selection of the number of groups is required; Infomap will continue to create sub-clusters as long as the links between part of a cluster are stronger than those with the rest of it. Consequently, the optimum number of groupings is extracted from the data rather than being decided in advance by the researcher.

The output of our method is a hierarchical clustering of articles by stories, with high-level groupings repeatedly split into smaller groups of stories and the final level being individual article nodes. The level of clustering varies – some top-level clusters will be large, some will be small and some will be a single article not detected as part of any given story.

\section{Demonstration and Validation of the Results}

We demonstrate the applicability of our methods on a corpus of 61,864 news articles collected over a three month period from a variety of different UK online news sources\footnote{The article-level texts, URLs and titles were collected by repeatedly crawling news source websites; articles are from the BBC, Mail, Express, Guardian, Mirror, and Sun.}. We conducted pairwise similarity calculations on all articles within a three day moving window, and then separated these articles into story groups using Infomap as described above.

The validation of the results proceeds in two stages. First, we tested the performance of the similarity matching algorithm by hand-coding a series of 6,764 article pairs from within a 3-day window; each article was marked ``related'' or ``not related'' depending on the researcher's judgment about whether the articles were part of the same story. 6,727 of the pairs were adjudged not related, with only 37 related. This sparseness accurately reflects the difficulties with supervised classification identified above: without extraordinary amounts of hand-coding, there is not enough data in the coding to accurately train a supervised classifier.

The results of this validation are presented in Table~\ref{tab:accuracy}, which gives statistics for both our keyword and BM25F approaches, and our ensemble classifier. As we are looking for very small amounts of related articles within a large corpus of unrelated ones, measurement of true accuracy requires assessment of both precision (proportion of items matched which are truly matching) and recall (proportion of all truly matching items which are identified) statistics. F\textsubscript{1} is the harmonic mean of these two results and is a good overall measure for general purposes. 

Overall, the pair-wise precision, recall, and F\textsubscript{1} results are good for all classifiers, especially considering the extremely high specificity needed to classify given the hugely unbalanced size of the classes (related/unrelated) used for the matching. The keyword approach has slightly higher precision than the BM25F approach, however the ensemble classifier outperforms them both in terms of precision. Note that the levels for the accuracy metric are, in themselves, much less spectacular than they appear: a classifier reporting ``not related'' to every pairing would achieve 99.45\% accuracy on this dataset. 

\begin{table}[ht]
\centering
\begin{tabular}{l r r r}
\toprule
Classifier:        & Keyword & BM25F & Ensemble \\
\midrule
Accuracy           & 0.999   & 0.999 & 0.999 \\
Precision          & 0.907   & 0.831 & 0.924 \\
Recall             & 0.831   & 0.831 & 0.831 \\
F\textsubscript{1} & 0.867   & 0.831 & 0.875 \\
\midrule
True Positive      & 49      & 49    & 49    \\
True Negative      & 20641   & 20636 & 20642 \\
False Positive     & 5       & 10    & 4     \\
False Negative     & 10      & 10    & 10    \\
\midrule
$N_{articles}$      & 252     & 252   & 252   \\
$N_{article pairs}$ & 20705 & 20705 & 20705 \\ 
\bottomrule    
\multicolumn{4}{l}{\footnotesize Note that, as the method limits matching to articles published} \\
\multicolumn{4}{l}{\footnotesize within a window of 3 days of each other,} \\
\multicolumn{4}{l}{\footnotesize $N_{article pairs} \neq (N_{articles} \cdot (N_{articles} -1) \cdot 0.5)$.} \\
\end{tabular}
\caption{Validation results for the simple keyword, BM25F and final ensemble classifiers}
\label{tab:accuracy}
\end{table}

\subsection{Overall clustering}

As a further validation of our technique, we compare some of the descriptive results we produce with hypotheses and other descriptive results produced by the limited set of manually authored papers on the subject. Existing research on the subject, especially the ``media storm'' research we have described above, has made three basic propositions about news stories. First, in terms of prevalence, Harcup and O'Neill have found around 30\% of news articles are follow up pieces to an original \citep[p.1477]{Harcup2001}. Though they do not specify how many original articles attract a follow-up, the fact that there are so many follow-ups suggests that at least 50\% of the news publication agenda might be devoted to stories. In terms of ``mega-stories'', meanwhile, Boydstun et al. find that just over 11\% of news coverage is, on average, attributable to very large ``mega stories'', which they argue last for around 15 days on average \citep[p.520]{boydstun_two_2014}. Finally, several authors argue that the evolution of stories should be heavily right-skewed, with the majority of articles on the topic published shortly after the story breaks, whilst the volume of coverage then decays exponentially over time \citet[p.524]{vasterman_media-hype:_2005}. However, within this overall pattern, new smaller peaks may be observed as further developments emerge in the story \citep[p.197]{wien_anatomy_2009}.  

\begin{table}
\begin{center}
\begin{tabular}{l c c}
\toprule
Number of Articles & 61,864 & \\
Articles associated with an article in the same paper & 30,985 & 50\% \\
Articles associated with article in another paper & 31,882 & 52\% \\
\bottomrule
\end{tabular}
\caption{Article sample: overview}
\label{tab:artoverview}
\end{center}
\end{table}

Initial descriptive statistics for the stories and articles in our dataset are presented in Table~\ref{tab:artoverview}. In total there are 61,864 articles in our dataset. About 50\% of them are associated with at least one other article in the venue in which they are published; these articles are divided into 8,410 separate story clusters. This figure is quite similar to the one produced by Harcup and O'Neill. The distribution of the size of these clusters (see Table~\ref{tab:artclustersizes}) is heavy tailed, with the majority of these clusters having between 2 and 10 articles (and 4,412 with just 2 articles). These smaller clusters account for 40\% of articles in the total dataset. The larger clusters (with 11 articles or more) account in total for around 7\% of the dataset. This is broadly similar to \citeauthor{boydstun_two_2014}’s finding that 11\% of news coverage is attributable to very large stories. 

\begin{table}
\begin{center}
\begin{tabularx}{\linewidth}{X X X X X}
\toprule
Cluster size & N & Articles in these clusters & As \% of total articles & Average duration (days) \\
\midrule
2-10 & 8,063 & 24,961 & 40\% & 1.4 \\
11-20 & 271 & 3,746 & 6\% & 3.6 \\
21-30 & 45 & 1,129 & 2\% & 3.9 \\
31-40 & 25 & 863 & 1\% & 3.7 \\
40+ & 6 & 286 & 0\% & 4.9\\
\midrule
Total & 8,410 & 30,985 & 50\% & 1.4 \\
\bottomrule
\end{tabularx}
\caption{Article sample: cluster sizes}
\label{tab:artclustersizes}
\end{center}
\end{table}

\begin{figure}
\begin{center}
\includegraphics[width=\textwidth]{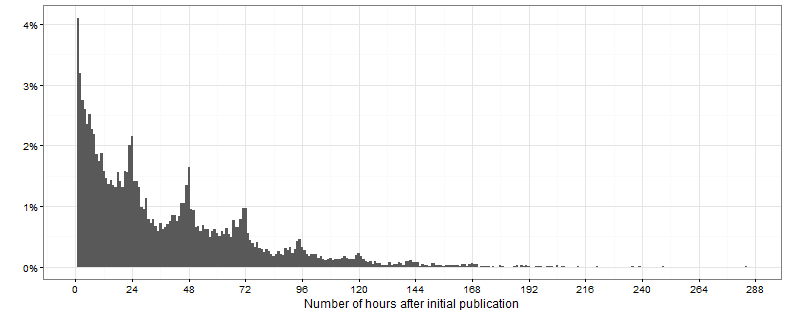}
\caption{Time between initial publication and follow-up articles}
\label{fig:timeafterpub}
\end{center}
\end{figure}

The average story in our dataset lasts 1.4 days. Larger stories are unsurprisingly longer ones as well, though once stories go above 11 articles in size there is not a great deal of difference between them: lasting on average around 3.5 – 5 days. Only one story in the dataset lasted for more than 10 days; no story lasts for the average of 15 days identified by \citeauthor{boydstun_two_2014}. The average evolution of large stories is shown in Figure~\ref{fig:timeafterpub}, which graphs the amount of time after an initial publication it took for new articles to appear for stories which reached 10 articles or more. We can see that, for example, just over 2\% of follow up stories appear exactly 24 hours after publication of the initial piece. The left skew and periodicity predicted by \citeauthor{wien_anatomy_2009} and \citeauthor{vasterman_media-hype:_2005} is clearly observable, with peaks roughly corresponding to daily news cycles. 

In other words, in a variety of respects our findings seem to confirm other descriptive results produced on the subject. The only area of disagreement concerns the length of the stories themselves, with our stories typically being smaller and shorter than ``mega stories'' identified in other research. Significantly, none of the stories identified by our technique were as long as the ones identified by Boydstun et al. 

In order to understand why, we conducted some manual inspection of the dataset and the resulting story clusters. During the period of data collection, there were two stories which dominated the news: the bombing of the Boston marathon and the murder of Lee Rigby, a British soldier in London. When we examined the clustering of news stories related to these topics, we found that the method had separated out these large stories into distinct ``sub stories'', each one relating to a different facet of the incidents themselves. This explains why the size of the stories themselves was smaller than hypothesised.

The reason for this sub-clustering is that the strength of the links between the smaller groups within the story are stronger than those between random articles from within the story as a whole. The Infomap algorithm, using information theoretic principles to extract structure, assigns articles to sub-clusters which reflect this. This raises a number of theoretical questions about the proper scope of a given story, questions which cannot be answered automatically. In the Boston case, the articles assigned to the sub-stories identified (e.g. coverage of the immediate aftermath of the bombing being separate from coverage of the arraignment) really do have more in common with each other than with the Boston story as a general whole. Whether ``Boston'' is the desired unit of analysis, or the more granular parts of it are, is a matter for the researcher. Interestingly however, the use of network methods to create our original clustering also offers the opportunity to partially resolve this situation, by indicating not only relations between articles but also relations between story clusters.  

\section{Conclusion}

The grouping of articles into clusters of related stories by computational methods is an interesting open research problem, of significant value to communications researchers studying news output at scale and particularly in relation to the study of media storms. This paper has introduced a new method, based on information retrieval tools, which gives researchers the opportunity to do computational work using the news story as a unit of analysis in addition to the article. Obvious applications of this approach are in analyses of news sources’ choices of stories to cover and the nature of that coverage, but there are many other questions which would benefit from the ability to handle grouped article data.

More broadly, although we have concentrated on the development of a model for grouping articles into news stories, the use of Information Retrieval tools potentially goes much wider. The underlying approach could be applied to many communications problems for which there is a desire to work with the whole of a given set of textual outputs (rather than an arbitrary selection).

This method is in no sense the final word on the subject. It achieves good validated results on the first part of its process, and good subjective results on the second which are consistent with previous work on media storms. But it's also clear that there remains much scope for further work: a large number of possible approaches could be taken to the pairwise matching step, and there are several network partitioning approaches that could be applied.

In terms of future research, there are many possible directions to take. Our method differs from supervised classification approaches in that the results depend on the quality of the pairwise similarity measurement rather than the quality of initial input coding. More could be done to improve precision and recall measures for this stage of the process, by drawing from an extensive information retrieval literature. As the second part of the process, partitioning the resulting similarity network, is agnostic as to the method for measuring similarity, the basic approach would continue to carry forward.

If a particular piece of research is aided most by maximising precision at the expense of recall (because you need all identified items to be similar, but don't need to identify all similar items) then a simple manual analysis of the identified links could increase precision to arbitrarily high levels. In this case, the information retrieval tools would be acting as an automated filter, running through the huge number of possible links and identifying a tiny percentage for manual analysis. Although not fully automated, this approach is perfectly valid in many situations\footnote{It also reflects how search engines use information retrieval techniques in the real world. Google needs to provide a great result on the front page, but if only 9 out of 10 of its links are to suitable sources then the human user will simply ignore those which are obviously wrong.}.

More broadly, the study of news at scale would benefit both from the exploration of other approaches to article clustering, and also from the application of this paper's method to substantive research questions. We would advocate for both.

\printbibliography

\end{document}